# THEORY OF NEW QUANTUM OSCILLATIONS IN THE MAGNETORESISTANCE OF GRAPHENE LAYERS.

*N. García*
*Laboratorio de Física de Sistemas Pequeños y Nanotecnología*
*Consejo Superior de Investigaciones Científicas,*
*Serrano 144, Madrid 28006 (Spain)*

## ABSTRACT

We present a theory presenting new quantum oscillations in the magnetoresistance that are revealed as fine structures superimposed to the Schubnikov-de-Haas oscillations. They may be observed in experiments on graphene layers as fine structures that until now seem to have been overseen or considered to be noise. These oscillations appear also in the behaviour of the resistance as a function of the gate voltage that changes the number of carriers or Fermi level. Experimental studies of these resonances should give information of the uniformity and defects of the samples and represent a new fine structure spectroscopy. Also the lateral sample size and quantum effects may explain the absence of magnetoresistance in a few grapheme layers. Experiments are proposed.

The study of graphite and graphitic systems has recently seen a fibril activity with the development of new techniques enabling the fabrication of samples of micron size and a few layers of grapheme (*FLG*). Quantum Hall effect (*QHE*) was observed in macroscopic graphite [1]. The report that by micromechanical manipulation one can obtain samples consisting of *FLG* of micron size [2] has set in action a large number of researches to understand the magnetoresistance (*MR*) properties of these microsystems that show changes in electronic properties and number of carriers of the material by applying an external gate voltage. The samples are basically 2D systems [3]. *QHE* measurements have lead to the discovery a new Berry phase in the magnetotransport that leads to a new *QHE* [4]. This is assigned to the presence of Dirac, mass less, electrons [5,6] that dominate the carriers in one single layer of graphene. When many layers are coupled to have graphite the carriers contain not only Dirac but also normal electrons. Then in the *MR* and *QHE* one can observe contributions of both types of carriers [7].

Electronic transport in microstructures of a *FLG* has shown to have very high mobility and minimum in conductance of $\frac{4e^2}{\hbar}$ at zero applied field [2,3,4,8,9] and *e* and *h* are the electron charge and the Planck constant. Typical resistances of the FLG are of the order of the inverse of this minimum (~6400Ω). This suggest clearly that in micronsize samples the transport is quantum mechanical and that the coherence length as well as the Fermi wavelength, *l and* $\lambda$ of the carriers are of the same order or larger than the sample size. Therefore, in this case, there should be a lot of size effects and subtle quantum mechanical effects that could help to better understand new physics that may have relevant technological applications, although the later remains to be seen. The classical Boltzmann´s transport in this case should be reconsidered.

Another observation is that in the samples there are regions of micronsize having different number of layers [3,4]. That is to say there maybe microregions of only with one layer and then other regions with two or three layers, etc. Also because we have a *2D* crystal, this cannot have long-range order and there should be regions that are ordered together with regions of disorder where the strains and tensions are released. This permits the existence of diffraction but not long range order [10]. Then it is clear that in micronsize *FGL* there are potential fluctuations. For example, assume that we have a region with 2 layers that continues in a region with one layer, then the potential of 1 layer is different that in two layers because of coupling between layers that is of the order of *40meV (see Fig.1a)*. These potential variations are of extreme importance in small size samples exhibiting quantum transport and should introduce resonances and extra structure in the *MR;i.e.* quantum mechanical interferences.

The aim of this paper is to show that the potential fluctuations in quantum transport give quantum mechanical oscillation resonances in the *MR* as well as in the resistance as a function of the applied voltage; i.e number of carriers or for the matter of fact *Fermi energy*. Indeed these resonant oscillations may have been seen in previous experiments on *FLG* (2,4) although they have been overseen or may have been considered as noise, but our claim is that this is not the case and are structures that provide much information on the perfection of the samples. They can be used as a kind of a fine structure spectroscopy superimposed over the Schubnikov-de-Haas (*SdH*) oscillations. Also we propose that the absence of MR observed in the experiments, at low, field is due to the small lateral size of the samples, a few microns, and to to the FGL. Much thicker graphite also manifests the same behaviour [11].

We are proposing a quantum mechanical contribution to the *MR* due to coherent transport in *2D*-like system exhibiting potential fluctuations in different positions of the sample. To illustrate it, we will present our theory for some examples that are solvable in a simple way but without missing generality. Consider first the resistance of a *2D* quantum mechanical system that exhibit a fluctuation of the potential energy of value $V_o$. For example assume as in *Fig.1a* that we have the current between two planes of graphene regions connected by one single layer. Then the potential will exhibit a potential fluctuation as that of *Fig.1b or 1c* depending if it is positive or negative the relative potential between one and two layers. Our problem is to know its resistance as a function of the Fermi energy (*E*) indicated by the dashed-dotted line. Then it is well known that the conductance or transmitivity, the inverse of the resistance, exhibits quantum mechanical resonances. In fact the transmitivity for electrons forming an angle $\alpha$ with the interface of energy perpendicular to interface of one and two planes, $E(\alpha) = E\cos^2(\alpha)$ can be obtained easily to be:

$$T(E,\alpha) = \frac{4(E(\alpha)-V_0)E(\alpha)}{((4(E(\alpha)-V_0)E(\alpha)+V_0^2\sin^2((E(\alpha)-V_0)^{1/2}L))} \qquad (1).$$

where the Planck constant has been taken to be unity. The mass of the carriers is taken in units such that the length (L) of the fluctuation is taken in units of $\lambda$ divided by $2\pi$. Now when a gate voltage (*Vg*) is applied to the system we have a number of carriers that has been observed to be proportional to the voltage (*n(Vg)~Vg*). In a 2D, we also have that *n~E*. In *Fig.2a* and *b* we plot the values of *T(E,0).E* and $\frac{1}{T(E,0)E}$, the conductance and resistance for normal incidence of the electrons *α=0*. The *T(E,0)* is multiplied by *E* that represents the number of carriers because *E~n~Vg*. We see that there are quantum mechanical resonant oscillations that correspond with the continuum resonances of the potential of *Fig.1b and c*. The conductance is given by the integral in $\alpha$ multiplied by *E* ( the number of carriers)

$$\sigma(E) = E \int_0^{\pi/2} T(E,\alpha)d\alpha \qquad (2)$$

and the resistance $R(E) = \frac{1}{\sigma(E)}$. These are plotted in Fig. 2c and d for $V_0$ positive and negative. It is seen that even if the oscillations of *Fig.2a* are smeared out by the integration they are still present in the conductance and resistance. The resonant structure is more pronounced for positive than for negative potential and is noticeable that for *V>0* there is not current if $E<V_0$, or in fact there is a weak tunnel current. We have calculated these curves because there exist experimental data observing these oscillations. For example, in *Fig.3a* we show *Fig.2a* of *Ref.2* representing data for *FLG* for the resistance as a function of the applied voltage *Vg* ( remember that $E\sim n\sim Vg$) for the temperatures *5, 70* and *300K*. We notice that the curves at *70* and *300K* do not exhibit any fine structure, while the *5K* data show a clear oscillatory structure (*Fig.2c*). There is a curiosity in the plot of the data in ref. 2. It is a fact that the real interest for *QHE* is for low *T*, however in the inset of *Fig.3a* the experimentalists plot the conductivity (inverse of resistivity) for the data at *70K* that are smooth. However the most interesting curve is the one at *5K* and this shows clear oscillatory behaviour but they are obviated even if they contain really new and important physics. Probably the authors thought that the oscillations were "noise". Our claim is that they represent an important new spectroscopy data. We plot the data for 5K digitised in *Fig.3b* comparing with our calculations in *Fig.2b*. It is seen that the oscillations appear clearly as well as the steps in the conductance. Therefore we conclude that our calculations and ideas on the new oscillations is part of the real life. There is also the effect that the resistivity peak is shifted to $Vg\approx 36V$ that may indicate that $V_0>0$ because this also give a shift in setting the scale of observable conductance. Although this needs more experiments looking carefully at the aspects discussed here, because the presence of defects and impurities in the sample may induce also a shift.

Next step that we have to clear out is how the previous discussion can affect at all to the *SdH* oscillations?. The answer is subtle. In fact, the *FLG* are very resistive (*5000Ω*) for micronsize samples, then if currents (*I*) of a few *μA* are applied we have potential drops ($V_d=I.R$) of tens *mV* in some microns. This creates an extra potential consisting in the fluctuation plus a linear term (see *Fig.1b*). The solution of the Schroedinger´s equation with this linear term, even if $V_0$ is zero, will produce resonances and oscillations in resistance as a function of $Vg\sim E$. Similar resonances have been seen scanning tunnelling spectroscopy [12]. The solutions are combinations of Airy functions. However in first approximation the linear term can be taken to be a constant potential $V_d/2$ and therefore the potential acting over the carriers is $V_0+V_d/2$. Then in the case at hand we could also play with the resonant structure by changing the current as part of the spectroscopy. Now the reader can easily see why there is also a fine oscillatory structure in the *SdH* oscillations. Notice that in the from maximum to the minimum resistance of the *SdH* oscillation the *R* changes for these experiments of the order of *10000Ω* by changing the applied magnetic field *B* (see Fig *.4a* taken from *Fig.3*, *ref.4*). Then as discussed before the $V_d$ changes, at constant current, from approximately zero at the minimum to *10000Ω.I=10mV*, at *1μA*. These potential drops cannot be neglected when they fall in a few microns and for Fermi energies of the same order. Then the changing of potential due to the variation of the applied field *B* can be written as:

$$V_d(B) = IAR_{SdH}(B,T)\frac{(1+\cos(\frac{2\pi B_f}{B}))}{2} \qquad (4)$$

$B_f$ is the frequency of the *SdH* oscillation. For each filed *B* the potential drop takes place in the sample length. Then there will be a new $V_{0eff}(B) = V_0 + \frac{V_d(B)}{2}$, where we have taken the above mentioned approximation of the linear field that is taken into account by the factor ½ dividing $V_d(B)$. By inserting this potential into formulas *(1)* and *(2)* we obtain the magnetoconcductance and magnetoresistance as a function of *B*. In *Fig.3b* are depicted these calculations together with the experimental data at the lower *T=1.7K*. The curves, calculated keeping now *E* constant and experiments, show clear fine structure oscillations overimposed over the *SdH* large oscillations. *B* has been taken to be $B_f=1$ at the maximum and *B=2* at the minimum, so that the argument in (4) changes in π. This fine structure is observed in the experiments up to *30K* and then for higher values of *T* they are more smeared out. The calculations in agreement with experiments show oscillations where the *SdH* resistance changes fast because in this region changes also fast. However at the maximum and minimum where there are weak changes of the potential not fine structure oscillation is observed. Fine structure should also manifest in the *QHE* in the transition regions between plateaus. The experiments clearly show also this behaviour; see *Fig.2 ref.4* for the fine structure oscillations in the transversal *MR*, *Rxy*, between *6* and *7 Tesla* that is a transition region between plateaus. In Fig.4c we plot the experimental data and our calculations taking into account the quantum oscillatory contributions. Both experiments and theory present a quite reasonable agreement although no tentative have been done to fit the data, but the qualitative behaviour is the same for theory and experiments.

We notice that in *Ref.2* the observation of the fine structure oscillations is weak but this is due to the fact that are measured at values of *Vg* where the changes of the *SdH* resistance changes a factor of *10* smaller than for those voltages used in *Ref.4*. It is manifest to us that the fine structure oscillations should be present by choosing adequately the parameters *B*, *T* and *Vg*. The fact that up to now the fine structures have been obviated is because they have not been realized before the new effect that the real physics shows and that has been introduced in this paper.

We have introduced quantum mechanical effects in the resistance as a function of *Vg* and in the *MR* superimposed to classical *SdH* oscillations. However the story is not finished because the coherence length of the electrons and the Fermi wavelength are of the order of microns, the sample size. In these conditions Boltzmann´s transport may not be a good approximation and all kind of finite size effects should appear. For example what is the meaning of the classical cyclotronic orbits then?. Notice that the classical radius is of the order of *0.05μ* and the Fermi wavelength of the electrons *1μ*. The Onsager quantization that gives rise to the Haas-van Halphen and *SdH* oscillations needs to be reconsidered because now the wavevector is not a continuum, etc. This may be a reason for explaining the absence of *MR*, thousand times smaller than for graphite [11], at low values of *B* observed on *FGL that* is not understood. In our opinion the sample size is plays an important role in reducing the MR. At this point we would like to propose some experiments in a FGL. One is that by increasing the lateral size of the sample the MR will increase. However and at the same time the resonance structure described in this work will reduce because the sample will have different fluctuations of different potentials and these smeared out the effect of just one fluctuation as described in this paper for a micron size sample. Different fluctuations in the same sample have similar effect that temperature and to show fine structure in a more random way because the resonance positions depend of the potential fluctuations. Also fluctuations like those sketched in Fig.1d should lead to localization phenomena.

There is a final question and this is: should the graphite show the effects discussed above because potential fluctuations have been observed by electric field microscopy [13]. The answer strictly speaking is yes. But for the reasons mentioned in the previous paragraph in macroscopic lateral size samples the effects tend to smeared out. Not only that, but increasing the thickness of the sample should have smearing effects on the resonant quantum oscillations. Then experiments reducing the size will reduce the MR but increase the strength of the quantum oscillations and the oscillations should be also observable

I thank M. Muñoz for helping me with the paper and discussions. Also I thank P. Esquinazi for discussions and introducing me to the experiments and F. Guinea for introducing me to the FLG experiments.

**FIGURE CAPTIONS.**

*Figure 1. a)* Sketch of a possible fluctuation, 2 layers of graphene regions are connected by 1 layer of graphene; b) potential profile between the points *A* and *B* for negative potential $V_0$, Fermi level is indicated by the dashed dotted lines .(c) the same as in (b) but with the dashed line indicating the variation of the applied potential (see text); (d) the same as in (b) but for positive potential, the dashed lines sketch possible additional fluctuations due to defects, impurities, etc

*Figure 2. a)* Transmittivity *T(E,0)* and its inverse (resistivity) for electrons with $\alpha=0$ for $V_0=-2$ as function of *E*. The value of *L=40* used correspond to a case where *E=meV* and the sample size is of the order of few microns. Notice the resonant structure.
b) The same for $V_0=2$. The transmitivity is only appreciable for $E>V_0$. c) Resistance and conductance *R(E)* and $T(E) = \dfrac{1}{R(E)}$ for $V_0=-2$. Notice the oscillations that although smeared out with respect to *R(E,0)* still exist. d) The same as in *c* but for $V_0=2$, now the T(E) is appreciable for $E>V_0$.

*Figure 3. a)* Experimental data, taken from *Fig 2* of *ref.2,* of the resistance, at 5K, for typical *FLG* as a function of *Vg (inset)*. Notice the fine oscillatory structure indicated by arrows. The Fig. shows the digitalised data of the inset where the resonances are clearly seen. Continuous line is our calculation in *Fig2.d*. The scale in abscissa has been rescaled considering *Vg~E*, see text.

*Figure 4. a)* Inset: data taken from *ref.4*, *Fig.3* for the resistance as a function of magnetic field *B* for different *T=1.7,4,10,20,30,40* and *60* approximately. The larger the variation of resistance the smaller the *T*. Notice the clear fine structure oscillations for *T<30K* that were obviated in *ref.4*. In the Figure we plot the data for *T=1.7K* together with our calculations for for $V_0=2$, *E=2.3*, $B_f=1$ and the *SdH* amplitude times the current (*I.A$R_{SdH}$(B,T)=2*), see *formula (4)* and text. The fine structure shows up in the calculations as well as in the experiments. In Fig.4b we present the experiment for the Hall resistance $R_{xy}$ taken from Fig.2a, ref.4 as a function of the field B and our calculations with the quantum oscillations. The arrows indicate the fine structure oscillations.

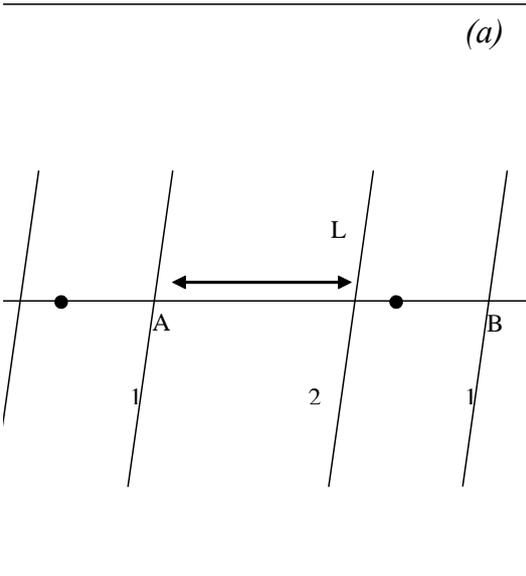
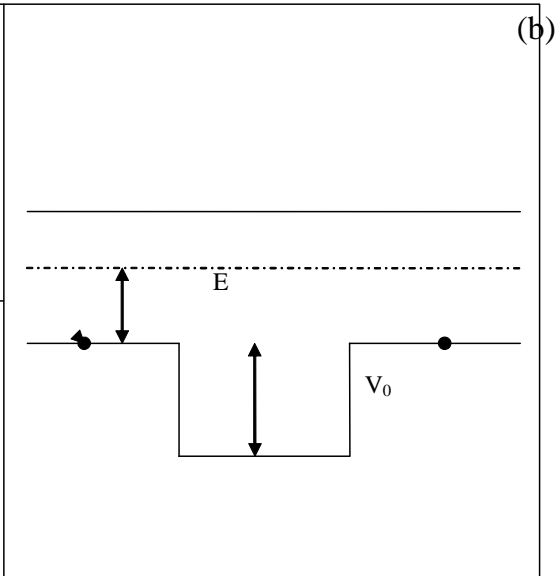
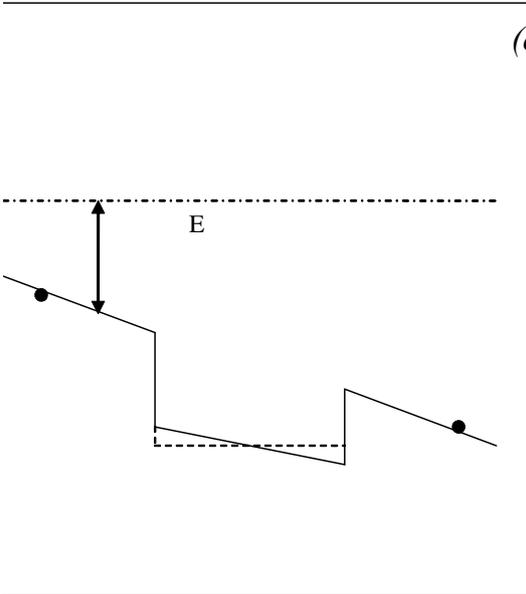
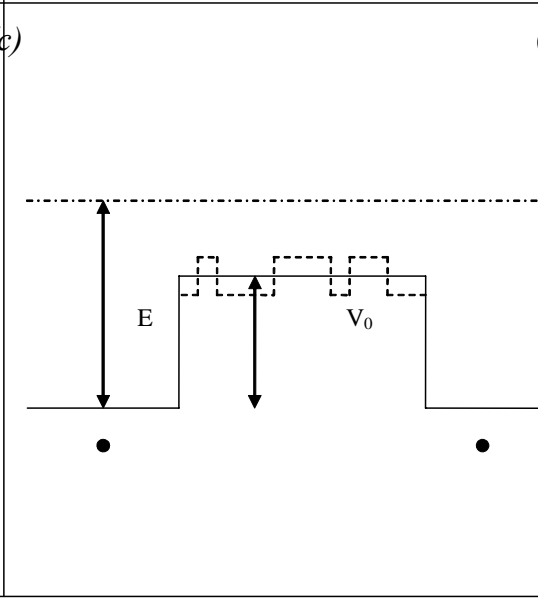

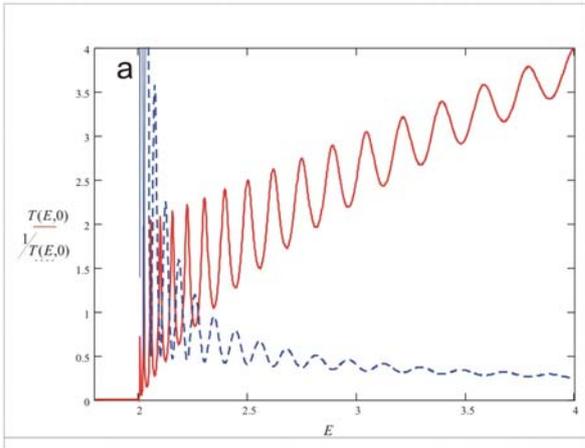
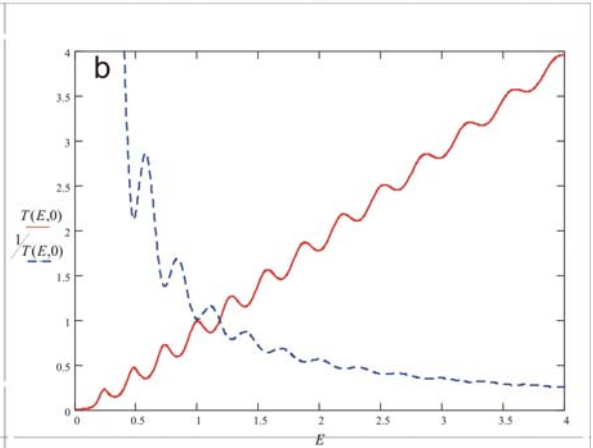
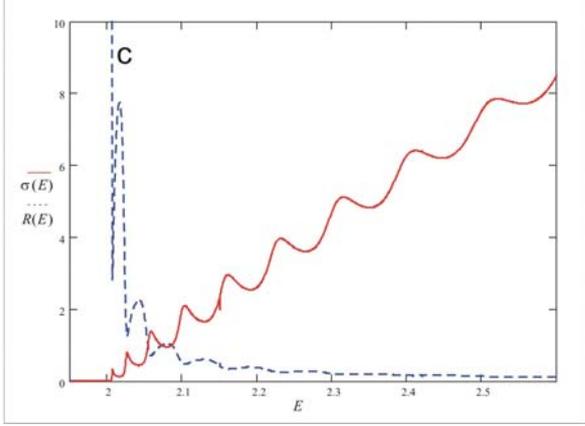
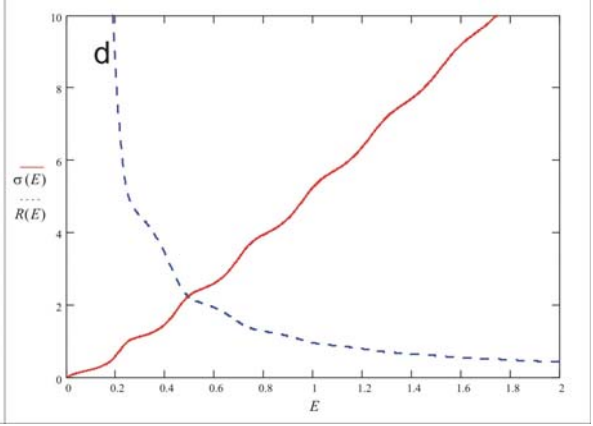

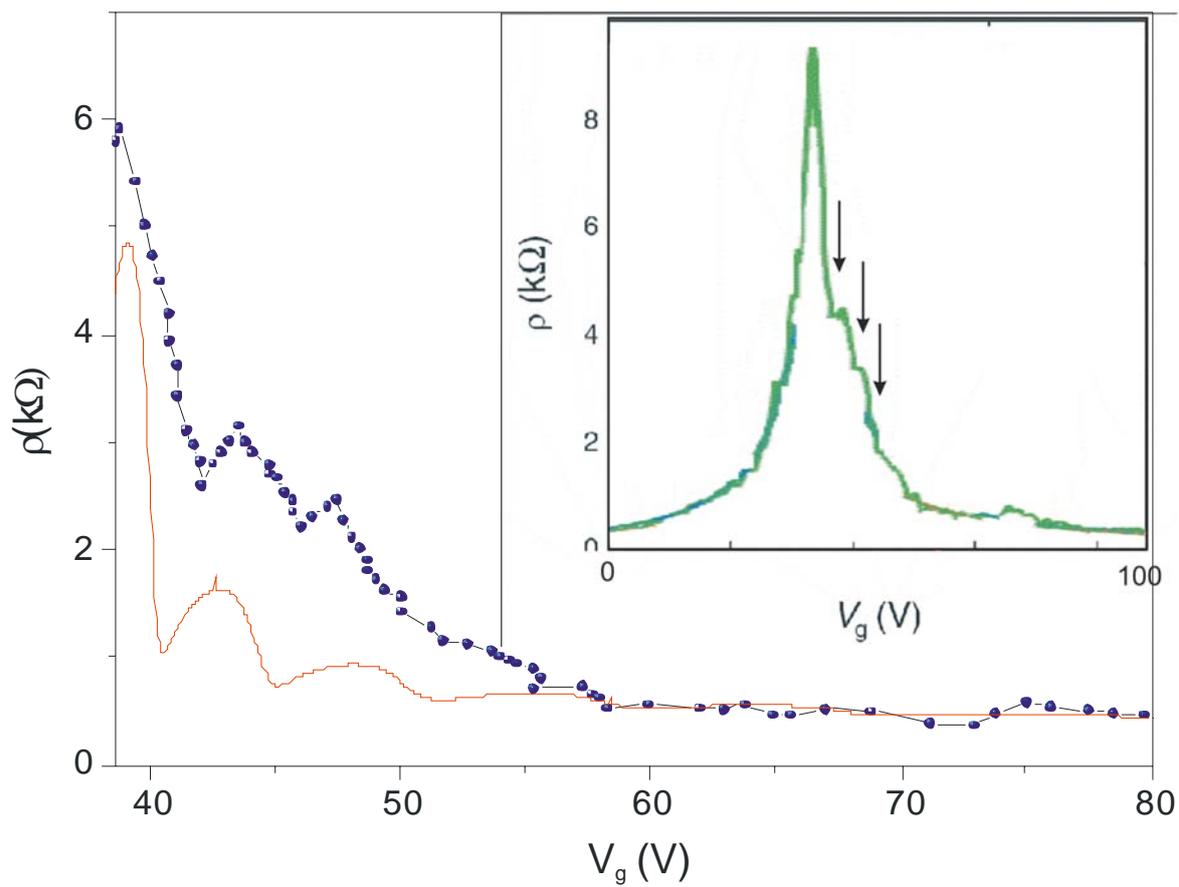

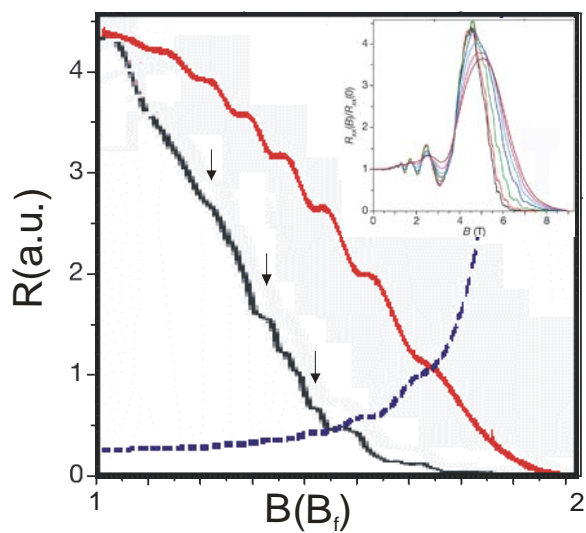 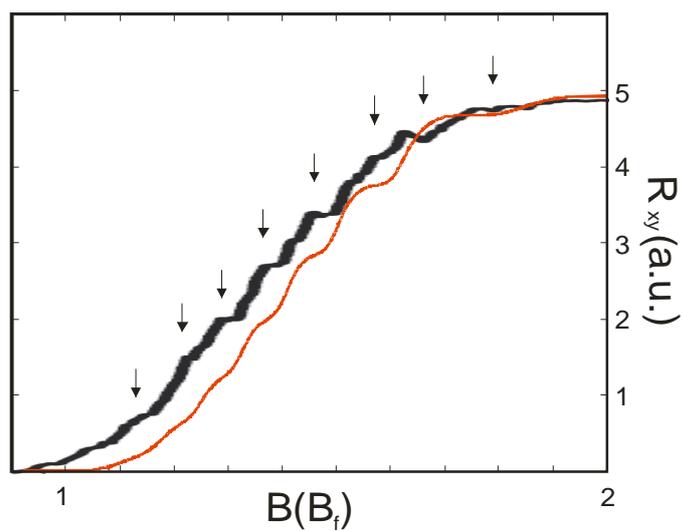